\documentclass[journal]{IEEEtran}

\usepackage{xparse}
\usepackage{tabularx}
\usepackage{comment}
\usepackage{epsfig,endnotes}
\usepackage{epsfig}
\usepackage{graphicx}
\usepackage{subfig}
\usepackage{cite}
\usepackage{url}
\usepackage{pifont}
\usepackage[dvipsnames]{xcolor}
\usepackage{wasysym}
\usepackage{threeparttable}
\usepackage{amssymb,amsfonts}
\usepackage{hyperref}
\usepackage{multirow}
\usepackage{subfig}
\usepackage{array}
\usepackage{listings}
\usepackage{booktabs}
\usepackage{makecell}
\usepackage{rotating}

\definecolor{myblue}{HTML}{1C52A3}

\hypersetup{colorlinks,urlcolor=black,linkcolor=myblue,anchorcolor=black,citecolor=myblue}

\newcolumntype{C}[1]{>{\centering\arraybackslash}p{#1}}
\newcolumntype{L}[1]{>{\raggedleft\arraybackslash}p{#1}}
\newcolumntype{R}[1]{>{\raggedright\arraybackslash}p{#1}}

\renewcommand{\paragraph}[1]{\vspace{0.02in}\noindent\textbf{\textit{#1}}}

\newcommand{\soa}{\vspace{0.02in}\noindent\textbf{{Literature Work. }}}
\newcommand{\ins}{\vspace{0.02in}\noindent\textbf{{Industrial Solutions. }}}
\newcommand{\rev}{\vspace{0.02in}\noindent\textbf{{Review and Analysis. }}}
\newcommand{\reo}{\vspace{0.02in}\noindent\textbf{{Research Opportunities. }}}

\begin{document}
\title{Securing Serverless Computing: Challenges, Solutions, and Opportunities}

\author{Xing~Li,~\IEEEmembership{Student Member,~IEEE,}
Xue~Leng, 
and~Yan~Chen,~\IEEEmembership{Fellow,~IEEE}

\thanks{X. Li and X. Leng are with the College of Computer Science and Technology, Zhejiang University, Hangzhou, 310027, China (e-mail: xing.li@zju.edu.cn; lengxue\_2015@outlook.com).}
\thanks{Y. Chen is with the Department of Computer Science, Northwestern University, Evanston, IL 60208, USA (e-mail: ychen@northwestern.edu).} 
\thanks{This work has been submitted to the IEEE for possible publication. Copyright may be transferred without notice, after which this version may no longer be accessible.}
}


\maketitle

\begin{abstract}
Serverless computing is a new cloud service model that reduces both cloud providers' and consumers' costs
through extremely agile development, operation, and charging mechanisms  and has been widely applied since its emergence.
Nevertheless, some characteristics of serverless computing, such as fragmented application boundaries, have raised new security challenges.
Considerable literature work has been committed to addressing these challenges. Commercial and open-source serverless platforms implement many security measures to enhance serverless environments.
This paper presents the first survey of serverless security that considers both literature work and industrial security measures. We summarize the primary security challenges, analyze corresponding solutions from the literature and industry, and identify potential research opportunities. Then, we conduct a gap analysis of the academic and industrial solutions as well as commercial and open-source serverless platforms' security capabilities, and finally, we present a complete picture of current serverless security research.

\end{abstract}


\IEEEpeerreviewmaketitle

\section{Introduction}
\label{sec:introduction}
\IEEEPARstart{T}{he} development of cloud computing has driven various service model innovations, including serverless computing. In this model, cloud providers are responsible for all server-related management tasks, such as resource allocation, service deployment, scaling, and monitoring, and an application is charged only for its execution time. As a result, consumers can avoid tedious management tasks, focus on the business code, and save costs by not paying for idle resources. These advantages in terms of efficiency and economy have led to the rapid development of serverless computing in recent years and have attracted extensive attention from both industry and academia.
According to a recent report, the global serverless market size is estimated to grow to \$21.1 billion by 2025 \cite{serverless_market}.


However, as a novel service model, serverless computing also has certain distinguishing features that present some challenges, leading to security and compliance concerns. For example, its agile and lightweight virtualization technologies may lead to weak isolation, and the ephemeral nature of its computing instances could increase the difficulty of security management.
In recent years, substantial efforts have been made to address these challenges. Related studies have arisen from academic research, commercial cloud providers, and thriving open-source communities and have considerably enhanced serverless security.
Therefore, a systematic survey of current research progress is needed to provide a foundation for continuing security enhancement.


Some previous surveys of serverless computing have been conducted \cite{jonas2019cloud, shafiei2020serverless}; however, they have three main drawbacks in revealing the current status of security research. In terms of subject matter, these surveys have extensively discussed the concepts, challenges, applications, and prospects of serverless computing but have not specifically targeted security. Regarding content, the previous surveys have briefly introduced possible risks or attacks but have not systematically analyzed existing solutions and future directions of research. In terms of materials, these surveys have focused on literature work and have not considered the measures adopted in industry (i.e., commercial or open-source platforms); however, an academic focus alone is insufficient for a review of a widely applied cloud computing model such as serverless computing.


Therefore, a complete and systematic review of serverless security is urgently needed. To promote serverless computing and inspire new research, we present the first survey on serverless security whose horizon is expanded from academia alone to include industry. In this paper, we first introduce the concept and background of serverless computing (\autoref{sec:background}). Starting from the unique characteristics of this service model, we then present a classification of the main security challenges and analyze their root causes. Subsequently, we review the state-of-the-art solutions proposed in the literature and adopted by popular serverless platforms for each challenge. Based on the degree to which these problems are solved, we note potential research opportunities (\autoref{sec:isolation}--\autoref{sec:confidentiality}).
Finally, we illustrate our findings in comparisons of the current academic and industrial solutions as well as commercial and open-source platforms (\autoref{sec:gap}).

The main contributions of this work are as follows:
\begin{itemize}
\item A systematic summary of the challenges arising in securing serverless computing is presented.
\item A brief but comprehensive review of academic and industrial solutions is provided.
\item A set of promising potential research opportunities is proposed.
\item A multiaspect gap analysis of the current research status is conducted.
\end{itemize}





\begin{figure*}[tb]
\centering
\includegraphics[width=2\columnwidth]{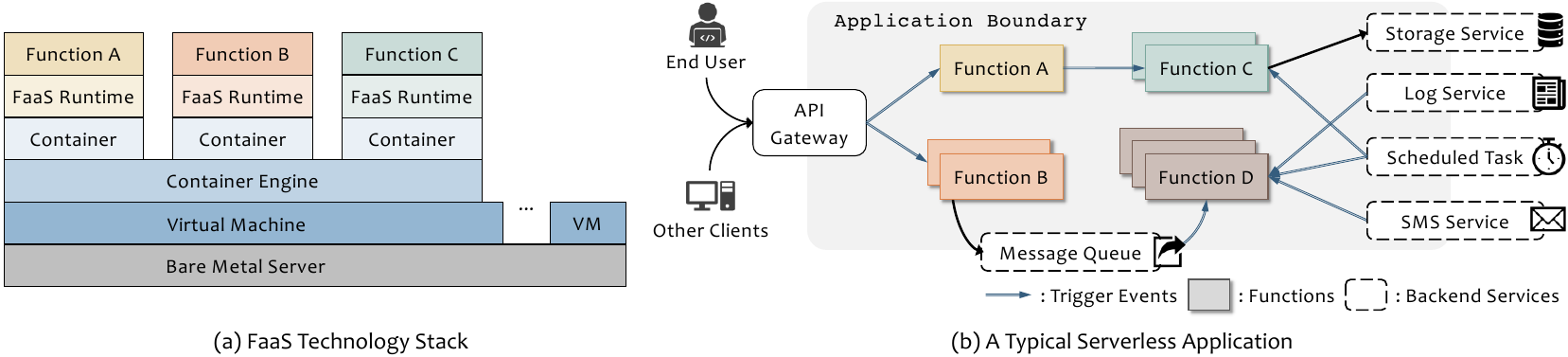}
\caption{(a) A container-based FaaS technology stack. Typically, only the top-level function code is managed by customers. (b) The architecture of a serverless application that includes FaaS and BaaS components.}
\label{fig:severless}
\end{figure*}

\section{Background and Overview}
\label{sec:background}

\subsection{Serverless Computing}

Virtualization technology is the cornerstone of cloud services.
Based on the \textit{platform-as-a-service} (PaaS) model, the recent prosperity of lightweight virtualization (e.g., \textit{containers}) has given birth to a new model: \textit{function-as-a-service} (FaaS).
As shown in \autoref{fig:severless} (a), in this model, customers develop and deploy small code pieces called \textit{functions} in the cloud. These functions are deployed in lightweight virtualized environments such as containers and executed on platform-provided runtimes. Each function focuses on a different task, and multiple functions can be combined in an event-driven manner to realize complex business logic. Events that can activate the execution of functions are called \textit{triggers}; these include but are not limited to HTTP requests, logs, storage events, and timers.



With lightweight virtualization, functions can be initialized extremely fast (usually within a few milliseconds). Therefore, providers can instantiate functions only when needed and flexibly scale them, thereby maximizing resource utilization. Consequently, functions usually have a short lifecycle.
This agility is also beneficial to consumers.
Since the occupied resources will be released when they are not in use, consumers need to pay only for functions' actual execution time and do not need to reserve resources for emergencies.

Moreover, cloud providers offer dedicated services and application programming interfaces (APIs) for tasks such as storage, logging, and identity management. These services can help customers build and manage server-side logic, thus significantly accelerating application development and release. This service model is known as \textit{backend-as-a-service} (BaaS).

\textit{Serverless computing} is generally regarded as a combination of FaaS and BaaS. By undertaking all management tasks directly related to infrastructure resources, cloud providers make server-related details transparent to consumers. From the customer's perspective, cloud applications are serverless --- neither development, deployment, management, nor billing is server-centric. \autoref{fig:severless} (b) illustrates the architecture of a typical serverless application.

Currently, serverless computing is widely applied. Major providers such as Amazon Web Services (AWS), Microsoft Azure, and Google Cloud have all launched serverless products; a series of open-source platforms is also emerging. Meanwhile, this new cloud service model has begun to be embraced in many scenarios, such as data processing and the Internet of Things (IoT) paradigm. 

\begin{table*}[tb]
\caption{The landscape of the survey, including security challenges and their root causes, proposed solutions, our reviews, and research opportunities.}
\label{tab:solution}
\footnotesize
\centering
\scalebox{0.993}{

\begin{tabularx}{\linewidth}{lXXXX}%
\toprule[1pt]
{\makecell[c]{\textbf{Challenges}}} &
  \makecell[c]{\textbf{Resource Isolation}} &
  \makecell[c]{\textbf{Security Monitoring}} &
  \makecell[c]{\textbf{Security Management}} &
  \makecell[c]{\textbf{Data Protection}} \\ \hline
  
\rule{0pt}{8pt}\multirow{3}{*}{\makecell[l]{\textbf{Root} \textbf{Causes}}} &
  {The weakness of lightweight virtualization technologies in isolation} &
  {The ephemerality of functions and the broken boundaries of serverless applications} &
  The distributed nature of functions and the fragmented application boundaries &
  Platforms' control of the function lifecycle and BaaS services' participation in business logic \\ \hline
   
\rule{0pt}{8pt}\multirow{4}{*}{\makecell[l]{\textbf{Literature}\\\textbf{Work}}} &
  {Virtual-machine-based secure containers, e.g., minimized VMMs \cite{agache2020firecracker} and unikernels \cite{cadden2020seuss} } &
  {Information flow mining and tracing, e.g., static analysis \cite{obetz2019static} and dynamic tracing \cite{lin2018tracking,datta2020valve} } &
  {Advanced management capabilities, e.g., workflow-sensitive authorization \cite{sankaran2020workflow} and secure container network stacks \cite{nam2020bastion}} &
  {Hardware-based trusted execution environments, e.g., protecting running code with Intel SGX \cite{qiang2018se,brenner2019trust} }\\ \hline
  
\rule{0pt}{8pt}\multirow{4}{*}{\makecell[l]{\textbf{Industrial} \\ \textbf{Solutions}}} &
  {Secure containers and network boundaries, e.g., Firecracker \cite{agache2020firecracker}, gVisor, and VPCs} &
  {Security scanning and monitoring tools, e.g., Azure Monitor, AWS X-Ray, and DevSecOps tools} &
  {Fine-grained authentication and authorization, e.g., IAM systems and role-based access control} &
  {Encryption in transit and at rest, e.g., TLS/SSL, code/data encryption, and key management services} \\ \hline
  
  \rule{0pt}{8pt}\multirow{1}{*}{\makecell[l]{\textbf{Status}}} &
  {\makecell[l]{Well resolved}} &
  {\makecell[l]{Partially resolved}} &
  {\makecell[l]{Partially resolved}} &
  {\makecell[l]{Partially resolved}} \\ \hline
  
\rule{0pt}{8pt}\multirow{3}{*}{\makecell[l]{\textbf{Research}\\ \textbf{Opportunities}}} &
  Secure containers with better isolation, better performance, and lower overhead &
  Nonintrusive tracing schemes and advanced insight capabilities such as diagnosis and forensics & 
  Automatic security configuration and auditing tools & 
  Other confidential computing solutions such as homomorphic encryption \\
  \bottomrule[1pt]
\end{tabularx}
}
\end{table*}


\subsection{Threats and Security Challenges}
\label{sec:threat}

\begin{figure}[tb]
\centering
\includegraphics[width=1\columnwidth]{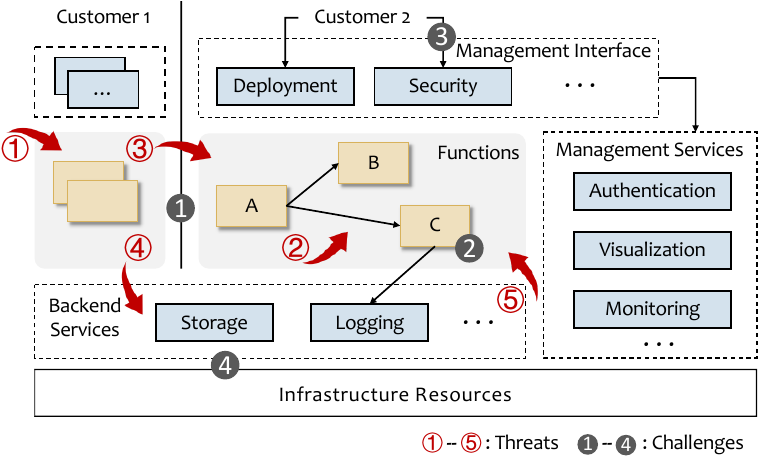}
\caption{Locations of threats and security challenges in the serverless computing framework. }
\label{fig:challenges}
\end{figure}

As a multitenant cloud service model, serverless computing is susceptible to security threats that can be divided into five categories based on where they are launched.
The first category consists of external attacks on applications from malicious users (\ding{172}), such as cross-site scripting attacks and injection attacks. Due to the unlimited scalability and the pay-as-you-go feature of serverless computing, denial of service attacks will lead to a substantial increase in the cost to application owners. The second comprises internal attacks on applications from malicious insiders (\ding{173}), such as illegal internal access and sniffing attacks. Adversaries in the internal network can even perceive sensitive information from the communication pattern and activity level of functions. For example, in a health monitoring application, a particular sequence of triggered functions may reflect a specific health condition of the patient \cite{shafiei2020serverless}. The remaining three categories are horizontal attacks between tenants (\ding{174}, e.g., side-channel attacks), vertical attacks on serverless infrastructures from malicious tenants (\ding{175}, e.g., container escape attacks), and vertical attacks on applications from malicious platforms (\ding{176}).

To address these threats, the best practice is to build a defense-in-depth system to protect serverless applications and platforms at multiple stages and levels. However, the flexibility and agility of serverless applications expose them to a series of security challenges in the process of achieving this goal. These challenges are concentrated in four aspects: resource isolation (\ding{182}), security monitoring (\ding{183}), security management (\ding{184}), and data protection (\ding{185}). \autoref{fig:challenges} depicts the locations of the threats and challenges in the serverless framework.


\subsection{Survey Methodology and Objects}
To fully understand the current state of the art in serverless security technologies, we collected and investigated the solutions proposed in research work and the security mechanisms adopted in practice for serverless platforms. An overview of this survey is presented in \autoref{tab:solution}.

For the literature work, we collected 77 papers related to serverless security that have been published since 2014 (when the serverless concept emerged). We selected nine recent research papers published in high-reputation conferences or journals to showcase the academic world's latest achievements.
Moreover, we collected the security features of popular serverless platforms from their documentation. These platforms include the leading commercial platforms AWS Lambda, Google Cloud Functions, Azure Functions, Alibaba Cloud Function Compute, and IBM Cloud Functions as well as the popular open-source serverless frameworks OpenFaaS, Kubeless, Knative, Fission, Apache OpenWhisk, and Nuclio.

\section{Challenge 1: Resource Isolation}
\label{sec:isolation}
Serverless computing relies on the dynamic release and reuse of software and hardware to improve resource utilization and reduce costs. Thus, there is a strong requirement for resource isolation, especially among multiple tenants.

However, this requirement conflicts with the weak isolation of lightweight virtualization, thereby giving rise to this challenge.
Serverless computing requires swift and on-demand function initialization. To this end, lightweight virtualization technologies have been widely adopted.
Unlike traditional virtual machines (VMs) with guest operating systems (OSs), containers share the host's OS kernel, and thus, their isolation is weak.
Malicious consumers could employ this weakness to influence or attack other consumers located on the same host (e.g., via side-channel attacks). For example, Wang \textit{et al.} \cite{wang2018peeking} found that AWS Lambda suffers from performance isolation issues, and Azure Functions has placement vulnerabilities that facilitate side-channel attacks.




\soa To enhance the isolation of lightweight virtualization technologies and address this challenge, published research focuses on VM-based solutions. Compared with containers, VMs can provide better security boundaries but require a longer time to initialize. Hence, researchers have attempted to make VMs more lightweight to meet the flexibility requirements of serverless computing.
The basic idea is to reduce the size of the guest kernels, that is, remove unnecessary modules, to build ``microVMs'', even keeping only the libraries that will be used \cite{agache2020firecracker,cadden2020seuss}. Moreover, Cadden \textit{et al.} \cite{cadden2020seuss} employed unikernel snapshots to further accelerate initialization, and Agache \textit{et al.} \cite{agache2020firecracker} minimized virtual machine monitors (VMMs) to reduce the trusted computing base. These microVMs have successfully reduced the initialization time to the millisecond level while providing VM-level isolation.


\ins Led by Amazon, Firecracker \cite{agache2020firecracker} has been applied in AWS Lambda. Moreover, all surveyed commercial serverless platforms except IBM Functions utilize their own \textit{secure container} products. These containers enable advanced isolation by employing various techniques, such as microVMs (e.g., Firecracker) and container sandboxes (e.g., gVisor). Commercial platforms also provide mechanisms for network-layer isolation, such as \textit{virtual private clouds} (VPCs).


\rev
Secure containers provide VM-level resource isolation and have become a standard mode of implementation in production environments. Therefore, we consider this challenge well resolved.
However, due to hardware sharing, these approaches are still susceptible to side-channel attacks. Moreover, trade-offs remain between security and performance. For example, gVisor sacrifices significant performance for security (in particular, a 216$\times$ reduction in file opening speed) \cite{young2019true}.
Additionally, many existing efforts employ snapshots \cite{ustiugov2021benchmarking} or even memory sharing \cite{cadden2020seuss} to reduce the start-up time as much as possible, which also degrades security to a certain extent.

\reo
Looking forward, serverless computing will embrace new secure containers if they offer progress in the following three dimensions: (1) security, achieving better isolation than VMs; (2) efficiency, providing performance closer to that of physical machines; and (3) lightweight computing, incurring lower overheads.

\section{Challenge 2: Security Monitoring}
\label{sec:detection}

A serverless platform generally maintains a set of runtimes for functions, including commonly used packages of the supported programming languages.
Moreover, to meet diverse development needs, most providers also support custom runtimes and even provide function marketplaces.
The security risks posed by third-party packages and functions necessitate reliable monitoring capabilities to detect anomalies and determine the security status.

Furthermore, in a serverless environment, the ephemeral nature of functions and the broken application boundaries present new challenges.
First, although the short-lived nature of functions reduces the attack surface, it also significantly narrows the window for developers to discover and diagnose problems.
Second, as shown in \autoref{fig:severless} (b), a serverless application may have multiple entry points due to a variety of triggers. These entry points, coupled with BaaS services, cause the boundaries of serverless applications to be fragmented and further increase the difficulty of monitoring.
Moreover, lacking control over the execution environments and BaaS services, consumers must rely on the provided interfaces for monitoring, which may lead to incomplete views of the running status. For example, Lin \textit{et al.} \cite{lin2018tracking} noted that AWS Lambda's monitoring granularity is relatively coarse, and the tracking paths are not continuous across BaaS services.

\soa To address this challenge, researchers have attempted to mine and track the information flows in serverless applications from both static and dynamic perspectives.
Obetz \textit{et al.} \cite{obetz2019static} proposed an augmented call graph that considers BaaS services and explained the feasibility of obtaining the graph through static analysis.
To dynamically track where and how data flow, Datta \textit{et al.} \cite{datta2020valve} placed an agent in each function's container to proxy its network requests. These agents generated and propagated labels for inbound and outbound traffic to record the flow of information and could also restrict functions' behavior based on preconfigured policies. In addition, Lin \textit{et al.} \cite{lin2018tracking} proposed GammaRay, a dynamic tracking tool that can capture the causality of invocations across BaaS services to build complete trace graphs.




\ins Commercial platforms provide various tools and interfaces to record, aggregate, display, and analyze monitoring data. Some can even serve as triggers for automatic function management. They also aim to gradually enable distributed tracing while ensuring adequate performance.
Moreover, to further mitigate the impact of nonsecure third-party components, providers recommend continuous security validation in continuous integration and continuous deployment (CI/CD) pipelines and offer a series of tools, such as static vulnerability scanning and automatic penetration testing, for this purpose.




\rev
The proposed dynamic methods solve the problem of the coarse-grained and incomplete nature of the existing monitoring schemes to a certain extent. For example, data-oriented tracing \cite{datta2020valve} can be integrated with traditional behavior-oriented mechanisms.
Moreover, since functions have relatively low internal complexity, static-analysis-based methods are feasible. These approaches could be an effective supplement to dynamic methods, which may introduce considerable runtime overhead.
Following best practices in the development phase can also mitigate the threat posed by nonsecure third-party components.
Unfortunately, the information that can be obtained through static analysis is limited. Current dynamic solutions all require code instrumentation or agent embedding to enable end-to-end monitoring and tracking, resulting in high compilation and runtime overheads \cite{datta2020valve,lin2018tracking}.
Hence, we consider this challenge only partially resolved.

\reo
The exploration of nonintrusive tracking schemes is a potential research opportunity.
Furthermore, advanced insight capabilities are lacking in current platforms.
These capabilities include assessing security status, predicting and diagnosing faults, and conducting detection and forensics for security events. With various triggers, serverless applications can naturally perform ``feedback regulation'' for self-management.
Meanwhile, functions' short lifecycles and massive logs impose high requirements on the speed of solutions. We believe this scenario represents a research opportunity where artificial intelligence (AI)-based methods might be beneficial.

\section{Challenge 3: Security Management}
\label{sec:management}


Security management is a critical means for administrators to enforce security intentions and protect applications. As described in \autoref{sec:threat}, the diversity of the possible attacks in serverless environments necessitates comprehensive security management capabilities that can provide protection at multiple levels.

However, the fragmented application boundaries of serverless applications increase the difficulty of security management. The distributed nature of functions further exacerbates this challenge.
On the one hand, functions must carefully inspect the input and output at all their entry points; failure to comply with this principle may lead to injection attacks. On the other hand, an external request may trigger communication among multiple functions and BaaS services, necessitating fine-grained authentication and authorization among functions and services. This task can be challenging for large-scale or complex applications.



\soa
Traditional authentication and authorization mechanisms have been well studied. Building on this work, researchers are currently attempting to provide advanced security management capabilities at new levels.
Sankaran \textit{et al.} \cite{sankaran2020workflow} built a workflow-sensitive authorization mechanism for serverless applications. It proactively verifies external requests' permissions for all functions involved in their workflows at the application entry point. Thus, applications can reject illegal requests as early as possible, thereby avoiding partial processing of illegal requests and reducing the potential attack surface.
Moreover, Nam \textit{et al.} \cite{nam2020bastion} revealed a data leakage risk in container networks and redesigned a secure network stack to protect intercontainer communication.
Specifically, the method verifies and filters out illegal requests based on the intercontainer dependencies and utilizes end-to-end direct forwarding to prevent traffic exposure. It even improves the container network's performance.


\ins Based on various security policies, cloud providers have formulated ``best practices'' for security management.
These policies act on multiple layers and multiple resources and have the potential to provide promising management capabilities.
For example, commercial platforms achieve well-defined identity authentication and fine-grained authorization with mature identity access management (IAM) systems.
AWS Lambda even provides role-based policies, resource-based policies, access control lists, and other management methods for consumers to choose among and flexibly combine.


\rev
Based on fine-grained security management policies, existing solutions provide multilevel protection for serverless applications. Providers also document best practices as guidance.
However, a gap may inevitably exist between actual situations and the recommended best practices.
First, both academic \cite{sankaran2020workflow,nam2020bastion} and industrial solutions require administrators to customize security policies in accordance with the application architecture and business logic. Although server management is the providers' responsibility, the burden placed on consumers for security management has not been relieved but instead has been made even heavier: fine-grained policies require attentive design and careful configuration, which is a time-consuming and error-prone process.
Second, providers' marketing emphasizes only cost-effectiveness and convenience, making it easy for consumers to ignore their responsibility to ensure security. As mentioned above, any violation of design principles and best practices may compromise applications' security.
In other words, current mechanisms do not reduce the complexity of security management. Applications' safety depends on the extent to which best practices are followed.
Therefore, we consider this challenge partially resolved.

\reo
At present, the assistance of automated tools is urgently needed. One possible research direction is automated security configuration. The ideal tool would be able to perceive an application's architecture and status, extract the consumer's security intentions, and provide support for policy configuration, maintenance, and updating.
Another research direction concerns automated security audits, for which there is a growing market. The ideal tool is expected to not only check whether security measures are deployed but also evaluate the quality of these measures based on an application's specific characteristics.
We believe that the security risks posed by incorrect configurations can be alleviated by advancing research in these two directions.

\section{Challenge 4: Data Protection}
\label{sec:confidentiality}
Data security is the lifeline of cloud services. Unlike in other models, serverless providers control the whole application lifecycle (e.g., the deployment, scaling, execution, and termination of functions) and the entire software stack.
Moreover, as part of the application and business logic, BaaS services can directly access and process data. All these characteristics require consumers to have deep trust in the provider.

Nevertheless, whether during the execution of code in an uncontrolled environment or the processing of sensitive data through the platform's services, privacy and compliance concerns may arise. To alleviate these concerns, platforms need to implement strict data protection measures to clearly and fully persuade consumers.
Given the flexibility requirements of serverless environments, this task could be challenging, but it is essential for expanding the application scenarios of serverless computing.



\soa
In addition to protecting data at rest or in transit through encryption, recent work has focused on protecting data in use.
Researchers have attempted to establish a \textit{trusted execution environment} (TEE) for functions to relieve customers of the need to fully trust providers \cite{qiang2018se, brenner2019trust}.
Specifically, several approaches employ Intel SGX, a hardware-based mechanism that uses encrypted memory to protect running code and data from privileged software, even the OS.
Moreover, Qiang \textit{et al.} \cite{qiang2018se} proposed a two-way sandbox, that is, a WebAssembly sandbox nested in an SGX sandbox, to prevent a platform from accessing sensitive data processed by functions while also protecting the platform from untrusted code.

\ins
On the basis of their mature storage services and security management experience, commercial platforms provide adequate support for encryption in transit and at rest. This protection covers the files and code uploaded by customers as well as the environment variables of functions. Moreover, standard key management services (KMS) are provided to help customers handle secrets.

\rev
Protecting data in use has become a significant dimension of cloud security.
In this regard, current serverless research is focused on hardware-based TEEs, which still have many limitations, such as poor performance, difficulty with hardware heterogeneity, and running only in the user space.
Moreover, these techniques have not yet been applied to serverless products. Therefore, we consider this challenge to be partially resolved.

\reo
In recent years, many confidential computing solutions, such as homomorphic encryption and secure multiparty computation, have emerged for the protection of data in the cloud.
Considering the limitations of the current solutions, the community is expected to appreciate further exploration and adoption of these techniques.

\section{Gap Analysis}
\label{sec:gap}

\renewcommand{\multirowsetup}{\centering}

\begin{table*}[tb]
    \caption{A brief list of security mechanisms used in commercial serverless platforms.}
\footnotesize
\centering
\scalebox{0.97}{
    \begin{tabular}{ll C{2.7cm}C{2.4cm}C{2.8cm}C{2.3cm}C{2.1cm}}
        \toprule[1pt]
        \multirow{2}{*}{\textbf{Aspect}} &
        \multirow{2}{*}{\textbf{Category}} & \multirow{2}{*}{\textbf{AWS Lambda}} & \textbf{Google} & \multirow{2}{*}{\textbf{Azure Functions}} & \textbf{Alibaba Cloud} & \textbf{IBM} \\
       & & & \textbf{Cloud Functions}& & \textbf{Function Compute} &\textbf{Cloud Functions}\\
        
        \hline
        \rule{0pt}{7.5pt}\multirow{3}{*}{\textbf{Isolation}}& \multirow{2}{*}{\textbf{System}}& \multirow{2}{*}{Firecracker} & \multirow{2}{*}{gVisor} & Azure Container Instances & Alibaba Cloud Sandbox & \multirow{2}{*}{Docker containers} \\\cline{2-7}
       \rule{0pt}{7.5pt}  & \textbf{Network} &VPC &VPC &Azure ASE &VPC &VPC  \\
        \hline
        
        \rule{0pt}{7.5pt}\multirow{2}{*}{\textbf{Monitoring}} & \multirow{1}{*}{\textbf{Visualization}} & Amazon CloudWatch & Stackdriver Logging & Azure Monitor & CloudMonitor & Sisdig \\ \cline{2-7}
        \rule{0pt}{7.5pt} & \multirow{1}{*}{\textbf{Tracing}} & AWS X-Ray & Cloud Trace & Application Insights & Context objects & Zipkin/Jaeger \\\hline
        
        \rule{0pt}{7.5pt}\multirow{5}{*}{\textbf{Management}} & \multirow{2}{*}{\textbf{Authentication}} & \multirow{2}{*}{AWS IAM} & {Cloud Functions IAM} & Azure AD and third-party identities & \multirow{2}{*}{Digital signatures} & \multirow{2}{*}{IBM IAM} \\ \cline{2-7}
        \rule{0pt}{7.5pt}& \multirow{3}{*}{\textbf{Authorization}} & Role-based/resource-based access control and other policies & \multirow{3}{2.4cm}{Role-based access control} & \multirow{3}{2.4cm}{Role-based access control} & \multirow{3}{2.4cm}{Role-based access control} & \multirow{3}{2.4cm}{Role-based access control} \\ \hline

        \rule{0pt}{7.5pt}\multirow{3}{*}{\textbf{Encryption}} & {\textbf{In Transit}} & mTLS & Google ALTS & mTLS & {TLS on ingress} & {TLS on ingress}\\ \cline{2-7}
        \rule{0pt}{7.5pt}& \multirow{2}{*}{\textbf{At Rest}} & Files, secrets, environment variables & \multirow{2}{*}{All data, KMS} & \multirow{2}{*}{All data, Key Vault} & Code, environment variables, KMS & \multirow{2}{*}{Data, secrets} \\  \hline

        \rule{0pt}{7.5pt}\multirow{2}{*}{\textbf{Tools}} & \multirow{2}{*}{\textbf{Tools}} & \multirow{2}{*}{AWS Trusted Advisor} & Event Threat Detection (ETD) & DevSecOps tools, Azure AD Access Reviews & \multirow{2}{*}{-} & \multirow{2}{*}{-} \\
        \bottomrule[1pt]
    \end{tabular}
    
    }
    \label{tab:platform}
\end{table*}

\begin{table*}[tb]
    \caption{A brief list of security mechanisms used in open-source serverless platforms.}
\footnotesize
\centering
\scalebox{0.97}{
    \begin{tabular}{ll C{1.8cm}C{2cm}C{2.1cm}C{2.2cm}C{2cm}C{1.8cm}}
        \toprule[1pt]
        {\textbf{Aspect}} & {\textbf{Category}} & {\textbf{OpenFaaS}} & {\textbf{Kubeless}} & {\textbf{Knative}} & {\textbf{Fission}} & {\textbf{OpenWhisk}}  & {\textbf{Nuclio}} \\

        \hline
        
        \rule{0pt}{7.5pt}\multirow{2}{*}{\textbf{Isolation}}& \textbf{System} & Docker & Docker & Docker & {Docker} & {Docker} & Docker \\ \cline{2-8}
        \rule{0pt}{7.5pt}& \multirow{1}{*}{\textbf{Network}} & K8s namespace & K8s namespace & K8s namespace & K8s namespace & K8s namespace & K8s namespace  \\ \hline
        
        \rule{0pt}{7.5pt}\multirow{2}{*}{\textbf{Monitoring}} & {\textbf{Visualization}} & Grafana & Grafana & Grafana & Grafana & Kamon & Dashboard \\ \cline{2-8}
        \rule{0pt}{7.5pt}& {\textbf{Tracing}} & - & - & Zipkin/Jaeger & Zipkin/Jaeger & Zipkin/Jaeger & -  \\\hline
        
        \rule{0pt}{7.5pt}\multirow{3}{*}{\textbf{Management}} & \multirow{2}{*}{\textbf{Authentication}} & \multirow{2}{*}{OAuth2} & Kong authentication &  Istio authentication& \multirow{2}{*}{DID key method} & Self-hosted authentication  & \multirow{2}{*}{-} \\ \cline{2-8}
        \rule{0pt}{7.5pt}& \multirow{1}{*}{\textbf{Authorization}} & K8s RBAC & K8s RBAC & K8s/Istio RBAC & K8s/Istio RBAC & Support  & - \\ \hline
        
        \rule{0pt}{7.5pt}\multirow{2}{*}{\textbf{Encryption}} & \multirow{1}{*}{\textbf{In Transit}} & {TLS on ingress} &{TLS on ingress} & Istio mTLS & Istio mTLS &{TLS on ingress}&{TLS on ingress}\\ \cline{2-8}
        \rule{0pt}{7.5pt}& \multirow{1}{*}{\textbf{At Rest}} & K8s secrets & K8s secrets & K8s secrets & K8s secrets & K8s secrets & K8s secrets\\

        \bottomrule[1pt]
    \end{tabular}
    }
    \label{tab:opensource}
\end{table*}

During our investigation, we obtained some interesting findings about the current research status of serverless security, which may imply potential directions towards a better ecosystem.

\subsection{Academic Solutions vs. Industrial Solutions}
\label{com:academic}
We compared the efforts in academia and industry from three perspectives: research interests, features of the solutions, and application status.
As mentioned above, both academic and industrial solutions cover the four main security challenges. This consistency in research interests implies that the current research is in a healthy state.

Nonetheless, the solutions show different characteristics. Academic research tends to demonstrate the feasibility of using specific methods, especially new technologies, to solve particular problems.
In contrast, the goal of commercial platforms is to be as secure as possible while ensuring stability. Thus, providers prefer methods that have been proven in long-term practice and usually combine several methods to achieve multilayer protection. In addition, research-led solutions often adopt a more holistic approach, from developer to platform. Without control over other parties, providers can promote their solutions only through guidelines or best practices.

Regarding application status, some of the reported academic research work has not yet been tested or applied in industry, as it is relatively new. We believe that open-source efforts from both academia and industry, such as publicly available research, real-world datasets, and standardized benchmarks, can help promote smooth adoption of newly emerging solutions.

\subsection{Commercial Platforms vs. Open-Source Platforms}
\label{sec:platforms}

Various commercial and open-source serverless platforms have flourished thanks to the vigorous serverless computing market. During our investigation, we found considerable differences in security between these two kinds of platforms. We discuss and analyze this phenomenon in this section.
Specifically, we compare the security measures adopted in commercial and open-source platforms in terms of five aspects: isolation, monitoring, management, encryption, and tools. The detailed results are listed in \autoref{tab:platform} and \autoref{tab:opensource}.

First, most commercial platforms are equipped with self-developed secure containers and provide VPCs for network-level isolation. In terms of monitoring, they introduce their own specialized distributed tracing and monitoring products into serverless scenarios and connect them to unified visualization consoles. Through security policies and IAM systems, these platforms provide mature authentication and authorization capabilities and support data encryption via storage services. Furthermore, their long-term investment and technological accumulation allow commercial platforms to provide many tools to assist customers, such as configuration recommendation and risk detection.

In contrast, most open-source platforms are built atop Kubernetes (K8s). Therefore, standard Docker containers and K8s namespace-based network isolation have become a general technical solution. For monitoring, these platforms usually integrate third-party tools or employ their infrastructures' capabilities to avoid reinventing the wheel. Moreover, open-source platforms mainly rely on ingress gateways and infrastructures such as K8s and Istio to achieve security management and encryption in transit. The secret management task is also outsourced to the underlying K8s infrastructure.
Similar to the monitoring aspect, open-source platforms usually do not provide security tools natively but support the integration of third-party tools such as Jenkins.

In terms of the above five aspects, the security measures of commercial platforms are all stronger than those of open-source platforms in terms of completeness and richness. The latter's security capabilities mainly rely on infrastructure or third-party tools.
Two main reasons for this difference exist.
First, commercial platforms have richer cloud management experience and more capital and human resources, so they can leverage their existing security mechanisms and provide standardized features as BaaS services (e.g., IAM systems). Second, the development of open-source platforms is still relatively rudimentary. They focus more on realizing core features of serverless computing, such as agile application construction and automatic function scaling, rather than security features.

Mature and standardized open-source platforms can counter the vendor lock-in problem and stimulate community efforts and related work.
Nevertheless, open-source platforms still have a long way to go to compete with commercial platforms in terms of security.
This security lag also limits their application.
Hence, more effort is needed to develop and standardize the security mechanisms of open-source serverless platforms. Although some have taken the lead (e.g., OpenFaaS), the competition among open-source serverless platforms is still fierce, and there is no de facto standard. From this perspective, unremitting efforts in developing security measures and the establishment of unique security capabilities therewith may become the key to gaining the dominant position.

\section{Conclusion}
\label{sec:conclusion}

This paper has presented a survey of both literature research and industrial solutions to elucidate the research status of techniques for securing serverless computing. Starting from the characteristics of serverless environments, we explained the origins of four main security challenges: resource isolation, security monitoring, security management, and data protection. Based on a review of existing solutions, we then identified possible research directions. Finally, we compared academic and industrial solutions as well as current commercial and open-source serverless platforms and elaborated on our thoughts regarding promising directions for establishing a better research and application ecosystem.

\vspace{7mm}
\bibliographystyle{IEEEtran}
\bibliography{IEEEabrv,ref.bib}

\begin{thebibliography}{10}
\providecommand{\url}[1]{#1}
\csname url@samestyle\endcsname
\providecommand{\newblock}{\relax}
\providecommand{\bibinfo}[2]{#2}
\providecommand{\BIBentrySTDinterwordspacing}{\spaceskip=0pt\relax}
\providecommand{\BIBentryALTinterwordstretchfactor}{4}
\providecommand{\BIBentryALTinterwordspacing}{\spaceskip=\fontdimen2\font plus
\BIBentryALTinterwordstretchfactor\fontdimen3\font minus
  \fontdimen4\font\relax}
\providecommand{\BIBforeignlanguage}[2]{{%
\expandafter\ifx\csname l@#1\endcsname\relax
\typeout{** WARNING: IEEEtran.bst: No hyphenation pattern has been}%
\typeout{** loaded for the language `#1'. Using the pattern for}%
\typeout{** the default language instead.}%
\else
\language=\csname l@#1\endcsname
\fi
#2}}
\providecommand{\BIBdecl}{\relax}
\BIBdecl

\bibitem{serverless_market}
\BIBentryALTinterwordspacing
MarketsandMarkets, ``{Serverless Architecture Market Size, Share and Global
  Market Forecast to 2025},'' 2020, accessed on 2021-03-15. [Online].
  Available: \url{http://bit.ly/serverless_market}
\BIBentrySTDinterwordspacing

\bibitem{jonas2019cloud}
E.~Jonas, J.~Schleier-Smith, V.~Sreekanti, C.-C. Tsai, A.~Khandelwal, Q.~Pu,
  V.~Shankar, J.~Carreira, K.~Krauth, N.~Yadwadkar \emph{et~al.}, ``Cloud
  programming simplified: A berkeley view on serverless computing,''
  \emph{arXiv preprint arXiv:1902.03383}, 2019.

\bibitem{shafiei2020serverless}
\BIBentryALTinterwordspacing
H.~Shafiei, A.~Khonsari, and P.~Mousavi, ``Serverless computing: A survey of
  opportunities, challenges and applications,'' Jan 2020, accessed on
  2021-03-15. [Online]. Available: \url{engrxiv.org/u8xth}
\BIBentrySTDinterwordspacing

\bibitem{agache2020firecracker}
A.~Agache, M.~Brooker, A.~Iordache, A.~Liguori, R.~Neugebauer, P.~Piwonka, and
  D.-M. Popa, ``Firecracker: Lightweight virtualization for serverless
  applications,'' in \emph{17th USENIX Symposium on Networked Systems Design
  and Implementation (NSDI 20)}, 2020, pp. 419--434.

\bibitem{cadden2020seuss}
J.~Cadden, T.~Unger, Y.~Awad, H.~Dong, O.~Krieger, and J.~Appavoo, ``Seuss:
  skip redundant paths to make serverless fast,'' in \emph{Proceedings of the
  Fifteenth European Conference on Computer Systems}, 2020, pp. 1--15.

\bibitem{obetz2019static}
M.~Obetz, S.~Patterson, and A.~Milanova, ``Static call graph construction in
  aws lambda serverless applications,'' in \emph{11th USENIX Workshop on Hot
  Topics in Cloud Computing (HotCloud 19)}, 2019.

\bibitem{lin2018tracking}
W.-T. Lin, C.~Krintz, R.~Wolski, M.~Zhang, X.~Cai, T.~Li, and W.~Xu, ``Tracking
  causal order in aws lambda applications,'' in \emph{2018 IEEE International
  Conference on Cloud Engineering (IC2E)}.\hskip 1em plus 0.5em minus
  0.4em\relax IEEE, 2018, pp. 50--60.

\bibitem{datta2020valve}
P.~Datta, P.~Kumar, T.~Morris, M.~Grace, A.~Rahmati, and A.~Bates, ``Valve:
  Securing function workflows on serverless computing platforms,'' in
  \emph{Proceedings of The Web Conference 2020}, 2020, pp. 939--950.

\bibitem{sankaran2020workflow}
A.~Sankaran, P.~Datta, and A.~Bates, ``Workflow integration alleviates identity
  and access management in serverless computing,'' in \emph{Annual Computer
  Security Applications Conference}, 2020, pp. 496--509.

\bibitem{nam2020bastion}
J.~Nam, S.~Lee, H.~Seo, P.~Porras, V.~Yegneswaran, and S.~Shin, ``Bastion: A
  security enforcement network stack for container networks,'' in \emph{2020
  USENIX Annual Technical Conference (USENIX ATC 20)}, 2020, pp. 81--95.

\bibitem{qiang2018se}
W.~Qiang, Z.~Dong, and H.~Jin, ``Se-lambda: Securing privacy-sensitive
  serverless applications using sgx enclave,'' in \emph{International
  Conference on Security and Privacy in Communication Systems}.\hskip 1em plus
  0.5em minus 0.4em\relax Springer, 2018, pp. 451--470.

\bibitem{brenner2019trust}
S.~Brenner and R.~Kapitza, ``Trust more, serverless,'' in \emph{Proceedings of
  the 12th ACM International Conference on Systems and Storage}, 2019, pp.
  33--43.

\bibitem{wang2018peeking}
L.~Wang, M.~Li, Y.~Zhang, T.~Ristenpart, and M.~Swift, ``Peeking behind the
  curtains of serverless platforms,'' in \emph{2018 USENIX Annual Technical
  Conference (USENIX ATC 18)}, 2018, pp. 133--146.

\bibitem{young2019true}
E.~G. Young, P.~Zhu, T.~Caraza-Harter, A.~C. Arpaci-Dusseau, and R.~H.
  Arpaci-Dusseau, ``The true cost of containing: A gvisor case study,'' in
  \emph{11th USENIX Workshop on Hot Topics in Cloud Computing (HotCloud 19)},
  2019.

\bibitem{ustiugov2021benchmarking}
D.~Ustiugov, P.~Petrov, M.~Kogias, E.~Bugnion, and B.~Grot, ``Benchmarking,
  analysis, and optimization of serverless function snapshots,'' in
  \emph{Proceedings of the 26th ACM International Conference on Architectural
  Support for Programming Languages and Operating Systems}, 2021, pp. 559--572.

\end{thebibliography}

%

\begin{IEEEbiographynophoto}{Xing Li}
[S'19] received a B.E. degree in software engineering from Shandong University, Jinan, China, in 2016. He is currently pursuing a Ph.D. degree in cybersecurity at Zhejiang University, Hangzhou, China.
His research interests are focused on the security of cloud networking and applications.
\end{IEEEbiographynophoto}

\begin{IEEEbiographynophoto}{Xue Leng}
received a Ph.D. degree in computer science and technology from Zhejiang University, Hangzhou, China, in 2020. Her research interests include security enhancement in SDN and performance optimization of microservices and service meshes.
\end{IEEEbiographynophoto}
\begin{IEEEbiographynophoto}{Yan Chen}
[F'17] received a Ph.D. degree in computer science from the University of California at Berkeley, Berkeley, CA, USA, in 2003. He is currently a Professor with the Department of Computer Science, Northwestern University, Evanston, IL, USA. His research interests include network security, measurement,
and diagnosis for large-scale networks and distributed systems.
\end{IEEEbiographynophoto}




\end{document}